\documentclass[manyauthors,nocleardouble,COMPASS]{cernphprep}

\usepackage{bm}
\usepackage{color}

\usepackage{amssymb}
\usepackage{amsmath}
\usepackage{amsbsy}
\usepackage{times}
\usepackage{epsfig}
\usepackage{colordvi}
\usepackage{graphicx}
\usepackage{wrapfig,rotating}
\usepackage[numbers, square, comma, sort&compress]{natbib}
\usepackage{hyperref} 
\usepackage{multicol}
\usepackage{booktabs}
\usepackage{multirow}
\usepackage{lineno}
\RequirePackage[T1]{fontenc}
 
\makeatletter

\DeclareSymbolFont{letters}     {OML}{cmm}{m}{it}
\DeclareSymbolFont{symbols}     {OMS}{cmsy}{m}{n}
\DeclareSymbolFont{largesymbols}{OMX}{cmex}{m}{n}
\DeclareMathOperator{\sgn}{sgn}

\begin{document}

\begin{titlepage}

\PHnumber{2015--199}
\PHdate{\today}

\title{Interplay among transversity induced asymmetries in 
hadron leptoproduction}

\Collaboration{The COMPASS Collaboration}
\ShortAuthor{The COMPASS Collaboration}

\begin{abstract}
In the fragmentation of a transversely polarized quark several left-right
asymmetries are possible for the hadrons in the jet. When only one unpolarized
hadron is selected, it exhibits an azimuthal modulation known as Collins
effect. When a pair of oppositely charged hadrons is observed, three asymmetries
can be considered, a di-hadron asymmetry and two single hadron asymmetries. In
lepton deep inelastic scattering on transversely polarized nucleons all these
asymmetries are coupled with the transversity distribution.  From the high
statistics COMPASS data on oppositely charged hadron-pair production we have
investigated for the first time the dependence of these three asymmetries on the
difference of the azimuthal angles of the two hadrons.  The similarity of
transversity induced single and di-hadron asymmetries is discussed.  A
new analysis of the data allows to establish quantitative
relationships among them, providing for the first time
strong experimental indication that the underlying
fragmentation mechanisms are all driven by a common physical process.
\end{abstract}
\vfill
\Submitted{(to be submitted to Phys. Lett. B)}

\end{titlepage}

{\pagestyle{empty} 
%
%
\section*{The COMPASS Collaboration}
\label{app:collab}
\renewcommand\labelenumi{\textsuperscript{\theenumi}~}
\renewcommand\theenumi{\arabic{enumi}}
\begin{flushleft}
C.~Adolph\Irefn{erlangen},
R.~Akhunzyanov\Irefn{dubna}, 
M.G.~Alexeev\Irefn{turin_u},
G.D.~Alexeev\Irefn{dubna}, 
A.~Amoroso\Irefnn{turin_u}{turin_i},
V.~Andrieux\Irefn{saclay},
V.~Anosov\Irefn{dubna}, 
W.~Augustyniak\Irefn{warsaw},  
A.~Austregesilo\Irefn{munichtu},
C.D.R.~Azevedo\Irefn{aveiro}
% ...
\end{flushleft}
%
%
\begin{Authlist}
\item \Idef{turin_p}{University of Eastern Piedmont, 15100 Alessandria, Italy}
\item \Idef{aveiro}{University of Aveiro, Department of Physics, 3810-193 Aveiro, Portugal} 
\item \Idef{bochum}{Universit\"at Bochum, Institut f\"ur Experimentalphysik, 44780 Bochum, Germany\Arefs{l}\Arefs{s}}
\item \Idef{bonniskp}{Universit\"at Bonn, Helmholtz-Institut f\"ur  Strahlen- und Kernphysik, 53115 Bonn, Germany\Arefs{l}}
\item \Idef{bonnpi}{Universit\"at Bonn, Physikalisches Institut, 53115 Bonn, Germany\Arefs{l}}
\item \Idef{brno}{Institute of Scientific Instruments, AS CR, 61264 Brno, Czech Republic\Arefs{m}}
\item \Idef{calcutta}{Matrivani Institute of Experimental Research \& Education, Calcutta-700 030, India\Arefs{n}}
\item \Idef{dubna}{Joint Institute for Nuclear Research, 141980 Dubna, Moscow region, Russia\Arefs{o}}
\item \Idef{erlangen}{Universit\"at Erlangen--N\"urnberg, Physikalisches Institut, 91054 Erlangen, Germany\Arefs{l}}
\item \Idef{freiburg}{Universit\"at Freiburg, Physikalisches Institut, 79104 Freiburg, Germany\Arefs{l}\Arefs{s}}
\item \Idef{cern}{CERN, 1211 Geneva 23, Switzerland}
\item \Idef{liberec}{Technical University in Liberec, 46117 Liberec, Czech Republic\Arefs{m}}
\item \Idef{lisbon}{LIP, 1000-149 Lisbon, Portugal\Arefs{p}}
\item \Idef{mainz}{Universit\"at Mainz, Institut f\"ur Kernphysik, 55099 Mainz, Germany\Arefs{l}}
\item \Idef{miyazaki}{University of Miyazaki, Miyazaki 889-2192, Japan\Arefs{q}}
\item \Idef{moscowlpi}{Lebedev Physical Institute, 119991 Moscow, Russia}
\item \Idef{munichtu}{Technische Universit\"at M\"unchen, Physik Department, 85748 Garching, Germany\Arefs{l}\Arefs{r}}
\item \Idef{nagoya}{Nagoya University, 464 Nagoya, Japan\Arefs{q}}
\item \Idef{praguecu}{Charles University in Prague, Faculty of Mathematics and Physics, 18000 Prague, Czech Republic\Arefs{m}}
\item \Idef{praguectu}{Czech Technical University in Prague, 16636 Prague, Czech Republic\Arefs{m}}
\item \Idef{protvino}{State Scientific Center Institute for High Energy Physics of National Research Center `Kurchatov Institute', 142281 Protvino, Russia}
\item \Idef{saclay}{CEA IRFU/SPhN Saclay, 91191 Gif-sur-Yvette, France\Arefs{s}}
\item \Idef{taipei}{Academia Sinica, Institute of Physics, Taipei, 11529 Taiwan}
\item \Idef{telaviv}{Tel Aviv University, School of Physics and Astronomy, 69978 Tel Aviv, Israel\Arefs{t}}
\item \Idef{triest_u}{University of Trieste, Department of Physics, 34127 Trieste, Italy}
\item \Idef{triest_i}{Trieste Section of INFN, 34127 Trieste, Italy}
\item \Idef{triest_ictp}{Abdus Salam ICTP, 34151 Trieste, Italy}
\item \Idef{turin_u}{University of Turin, Department of Physics, 10125 Turin, Italy}
\item \Idef{turin_i}{Torino Section of INFN, 10125 Turin, Italy}
\item \Idef{illinois}{University of Illinois at Urbana-Champaign, Department of Physics, Urbana, IL 61801-3080, U.S.A.}   
\item \Idef{warsaw}{National Centre for Nuclear Research, 00-681 Warsaw, Poland\Arefs{u} }
\item \Idef{warsawu}{University of Warsaw, Faculty of Physics, 02-093 Warsaw, Poland\Arefs{u} }
\item \Idef{warsawtu}{Warsaw University of Technology, Institute of Radioelectronics, 00-665 Warsaw, Poland\Arefs{u} }
\item \Idef{yamagata}{Yamagata University, Yamagata, 992-8510 Japan\Arefs{q} }
\end{Authlist}
%
%
\vspace*{-\baselineskip}\renewcommand\theenumi{\alph{enumi}}
\begin{Authlist}
\item \Adef{a}{Also at Instituto Superior T\'ecnico, Universidade de Lisboa, Lisbon, Portugal}
\item \Adef{b}{Also at Department of Physics, Pusan National University, Busan 609-735, Republic of Korea and at Physics Department, Brookhaven National Laboratory, Upton, NY 11973, U.S.A. }
\item \Adef{c}{Supported by the DFG Research Training Group Programme 1102  ``Physics at Hadron Accelerators''}
\item \Adef{d}{Also at Chubu University, Kasugai, Aichi, 487-8501 Japan\Arefs{q}}
\item \Adef{e}{Also at KEK, 1-1 Oho, Tsukuba, Ibaraki, 305-0801 Japan}
\item \Adef{f}{Present address: Universit\"at Bonn, Helmholtz-Institut f\"ur Strahlen- und Kernphysik, 53115 Bonn, Germany}
\item \Adef{g}{Also at Moscow Institute of Physics and Technology, Moscow Region, 141700, Russia}
\item \Adef{h}{Present address: RWTH Aachen University, III. Physikalisches Institut, 52056 Aachen, Germany}
\item \Adef{i}{Present address: Uppsala University, Box 516, SE-75120 Uppsala, Sweden}
\item \Adef{l}{Supported by the German Bundesministerium f\"ur Bildung und Forschung}
\item \Adef{m}{Supported by Czech Republic MEYS Grant LG13031}
\item \Adef{n}{Supported by SAIL (CSR), Govt.\ of India}
\item \Adef{o}{Supported by CERN-RFBR Grant 12-02-91500}
\item \Adef{p}{\raggedright Supported by the Portuguese FCT - Funda\c{c}\~{a}o para a Ci\^{e}ncia e Tecnologia, COMPETE and QREN, Grants CERN/FP/109323/2009, CERN/FP/116376/2010 and CERN/FP/123600/2011}
\item \Adef{q}{Supported by the MEXT and the JSPS under the Grants No.18002006, No.20540299 and No.18540281; Daiko Foundation and Yamada Foundation}
\item \Adef{r}{Supported by the DFG cluster of excellence `Origin and Structure of the Universe' (www.universe-cluster.de)}
\item \Adef{s}{Supported by EU FP7 (HadronPhysics3, Grant Agreement number 283286)}
\item \Adef{t}{Supported by the Israel Science Foundation, founded by the Israel Academy of Sciences and Humanities}
\item \Adef{u}{Supported by the Polish NCN Grant DEC-2011/01/M/ST2/02350}
\item [{\makebox[2mm][l]{\textsuperscript{*}}}] Deceased
\end{Authlist}
}
\newpage

\section{Introduction}

The description of the partonic structure of the nucleon at leading twist in the
collinear case requires the knowledge of three parton distribution functions
(PDFs), the number, helicity and transversity functions.  Very much like the
helicity distribution, which gives the longitudinal polarization of a quark in a
longitudinally polarized nucleon, the transversity distribution gives the
transverse polarization of a quark in a transversely polarized nucleon. Its
first moment, the tensor charge, is a fundamental property of the nucleon.
While the number and the helicity PDFs can be obtained from cross-section
measurements of unpolarized or doubly polarized lepton-nucleon deeply inelastic
scattering (DIS), respectively, the transversity distribution is chiral-odd and
as such can be measured only if folded with another chiral-odd quantity.  As
suggested more than 20 years ago~\cite{Efremov:1992pe,Collins:1992kk}, it can be
accessed in semi-inclusive DIS (SIDIS) off transversely polarized nucleons from
a left-right asymmetry of the hadrons produced in the struck quark fragmentation
with respect to the plane defined by the quark momentum and spin directions.
Recently, both the HERMES and the COMPASS experiments have provided unambiguous
evidence that transversity is different from zero by measuring SIDIS off
transversely polarized protons~\cite{Barone:2010zz}.  Two different processes
have been addressed. In the first process, a target spin azimuthal asymmetry in
single-hadron production is measured, the so-called Collins
asymmetry~\cite{Collins:1992kk}.  It depends on the convolution of transversity
and a hadron transverse-momentum dependent chiral-odd fragmentation function
(FF), the Collins function, which describes the correlation between the hadron
transverse momentum and the transverse polarization of the fragmenting quark.
The second process is the production of two oppositely charged
hadrons~\cite{Efremov:1992pe,Collins:1993kq,Jaffe:1997hf,Bianconi:1999cd}.  In
this case the so-called di-hadron target spin azimuthal asymmetry originates
from the coupling of transversity to a di-hadron FF, also referred as
interference FF, in principle independent from the Collins function.  In both
cases, measurements of the corresponding azimuthal asymmetries of the hadrons
produced in $e^+ e^-$
annihilation~\cite{Seidl:2008xc,Vossen:2011fk,TheBABAR:2013yha} provided
independent information on the two types of FFs, allowing for first extractions
of transversity from the SIDIS and $e^+ e^-$
data~\cite{Anselmino:2013vqa,Bacchetta:2012ty,Martin:2014wua,Kang:2014zza}.

The high precision COMPASS measurements on transversely polarized
protons~\cite{Adolph:2012sn,Adolph:2014fjw} showed that in the $x$-Bjorken
region, where the Collins asymmetry is different from zero and sizable, the
positive and negative hadron asymmetries exhibit a mirror symmetry and the
di-hadron asymmetry is very close to and somewhat larger than the Collins
asymmetry for positive hadrons.  These facts have been interpreted as
experimental evidence of a close relationship between the Collins and the
di-hadron asymmetries, hinting at a common physics origin of the two
FFs~\cite{Adolph:2014fjw,Bradamante:2014bqa,Braun:2015baa,Bradamante:2015xva}, as
suggested in the $^3$P$_0$ recursive string fragmentation
model~\cite{Artru:1995bh,Artru:2012zz} and, for large invariant mass of the
hadron pair, in Ref.~\cite{Zhou:2011ba}.  The interpretation is also supported
by calculations with a specific Monte Carlo model~\cite{Matevosyan:2013eia}.

In order to better investigate the relationship between the 
Collins asymmetry and the
di-hadron asymmetry the correlations between the azimuthal angles of the final
state hadrons produced in the SIDIS process $\mu p \rightarrow \mu' h^+h^-X$
have been studied using the COMPASS data.  
These correlations play an important role in the
understanding of the hadronization mechanism and in so far have been studied
only in unpolarized SIDIS~\cite{Arneodo:1986yc}.  
In this article for the first
time the results for SIDIS off transversely polarized protons are presented.  
The investigation has proceeded through three major steps:
\begin{itemize}
\item[{\em i})]
the Collins asymmetries for positive and negative hadrons
have been compared with the corresponding asymmetries 
measured in the SIDIS process
$\mu p \rightarrow \mu' h^+h^-X$,  i.e. when in the final state at least 
two oppositely charged hadrons are detected (2h sample);
\item[{\em ii})]
using the 2h sample the  asymmetries of 
$h^+$ and $h^-$
have been measured and their relation has been 
investigated as function of $\Delta \phi$, 
the difference of the azimuthal angles of the two hadrons;
\item[{\em iii})]
the dihadron asymmetry has been measured as function of $\Delta \phi$
and, using a new general expression, compared with the 
$h^+$ and $h^-$ asymmetries.
The integrated  values of the three  asymmetries have also been compared.
\end{itemize}

\section{The COMPASS experiment and data selection}

COMPASS is a fixed-target experiment at the CERN SPS taking data since
2002~\cite{Abbon:2007pq}.  The present results have been extracted from the data
collected in 2010 with a 160 GeV/c $\mu^+$ beam and a transversely polarized
proton (NH$_3$) target, already used to measure the transverse spin
asymmetries~\cite{Adolph:2012sn,Adolph:2012sp,Adolph:2014fjw}.  They refer to
the 2h sample, i.e.\ SIDIS events in which at least one positive and one negative
hadron have been detected.  

\begin{figure}[tb]
\begin{center}
\includegraphics[width=0.60\textwidth]{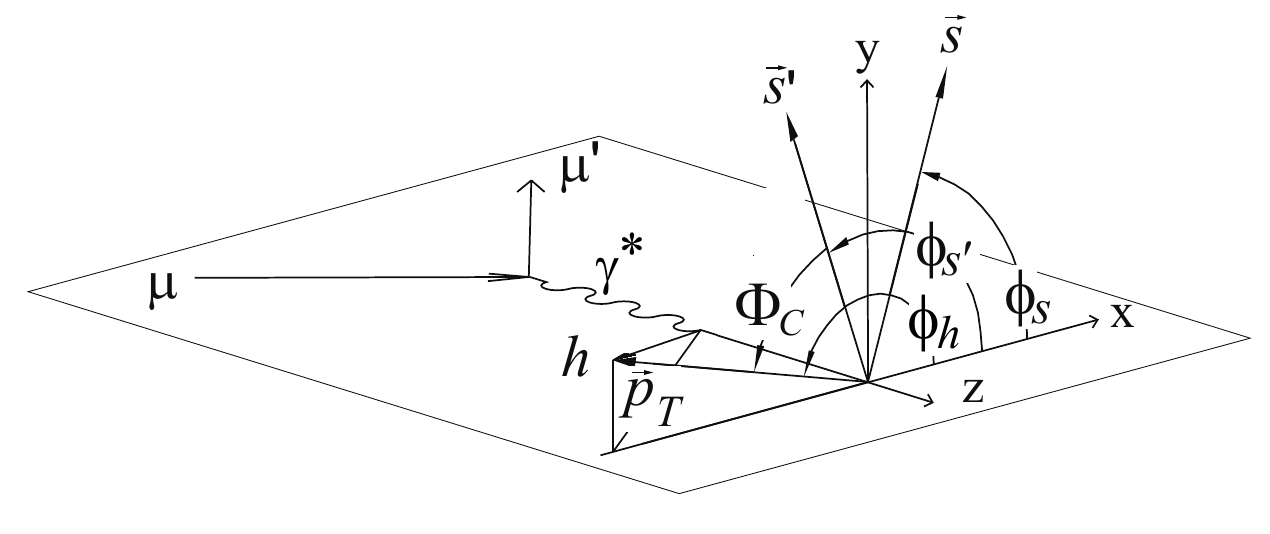}
\caption{Definition of the Collins angle $\Phi_C$ of a hadron h. 
The vectors $\vec{p}_T$, $\vec{s}$ and $\vec{s}^{\, \prime}$ are the
           hadron transverse momentum and the spin of the initial 
and struck quarks respectively.}
\label{fig:angles}
\end{center}
\end{figure}

The selection of the DIS events and of the hadrons
is described in detail in Ref.~\cite{Adolph:2014fjw}. 
 Standard cuts are applied
on the photon virtuality ($Q^2>1$ GeV$^2$/c$^2$), on the fractional energy
transfer to the virtual photon ($0.1<y<0.9$), and on the invariant mass of the
final hadronic state ($W>5$ GeV/c$^2$).  Specific to this analysis is the
requirement that each hadron must have a fraction of the virtual photon energy
$z_{1,2}>0.1$, where the subscript 1 refers to the positive hadron and subscript
2 to the negative hadron. A minimum value of 0.1 GeV/c for the hadron transverse
momenta $p_{T \,1,2}$ ensures good resolution in the azimuthal
angles.
As shown in Fig.~\ref{fig:angles}
the virtual photon direction is the z axis of the coordinate
  system while the x axis is directed along the lepton transverse
  momentum. 
The direction of the y axis is chosen to have a right-handed coordinate 
system.
Transverse components of vectors are defined with respect to z
  axis.

The 2h sample
consists of 33 million $h^+h^-$ pairs, to be compared with the 85 
million $h^+$ or 71 million
$h^-$ of the standard event sample (1h sample) of the previous
analysis~\cite{Adolph:2014fjw}, where at least one hadron (either positive or
negative)
per event was required. 

\section{Comparison of 1h and 2h sample asymmetries}

For each hadron the Collins angle $\Phi_{Ci}, \, i=1,2, \,$ is defined  as usual 
as $\Phi_{Ci}=\phi_i+\phi_S-\pi$, where $\phi_i$ is the azimuthal
angle of the hadron transverse momentum, $\phi_S$ is the azimuthal angle of the
transverse nucleon spin, and $\pi - \phi_S$ is the azimuthal angle of the spin
$\vec{s}^{\, \prime}$ of the struck quark~\cite{Kotzinian:1994dv}, as shown in
Fig.~\ref{fig:angles}. 
All the azimuthal angles
are measured around the $z$ axis. For the positive and
negative hadrons in the 2h sample, the amplitudes $A^{\sin\Phi_{C1}}_{CL1}$ and
$A^{\sin\Phi_{C2}}_{CL2}$ of the $\sin \Phi_{C1,2}$ modulations in the
cross-section have been extracted with the same method as
of Ref.~\cite{Adolph:2014fjw}
and labeled ``CL'' (Collins-like) to distinguish them from the standard Collins
asymmetries, which are defined in the 1h sample.
\begin{figure}[tb]
\begin{center}
\includegraphics[width=0.45\textwidth]{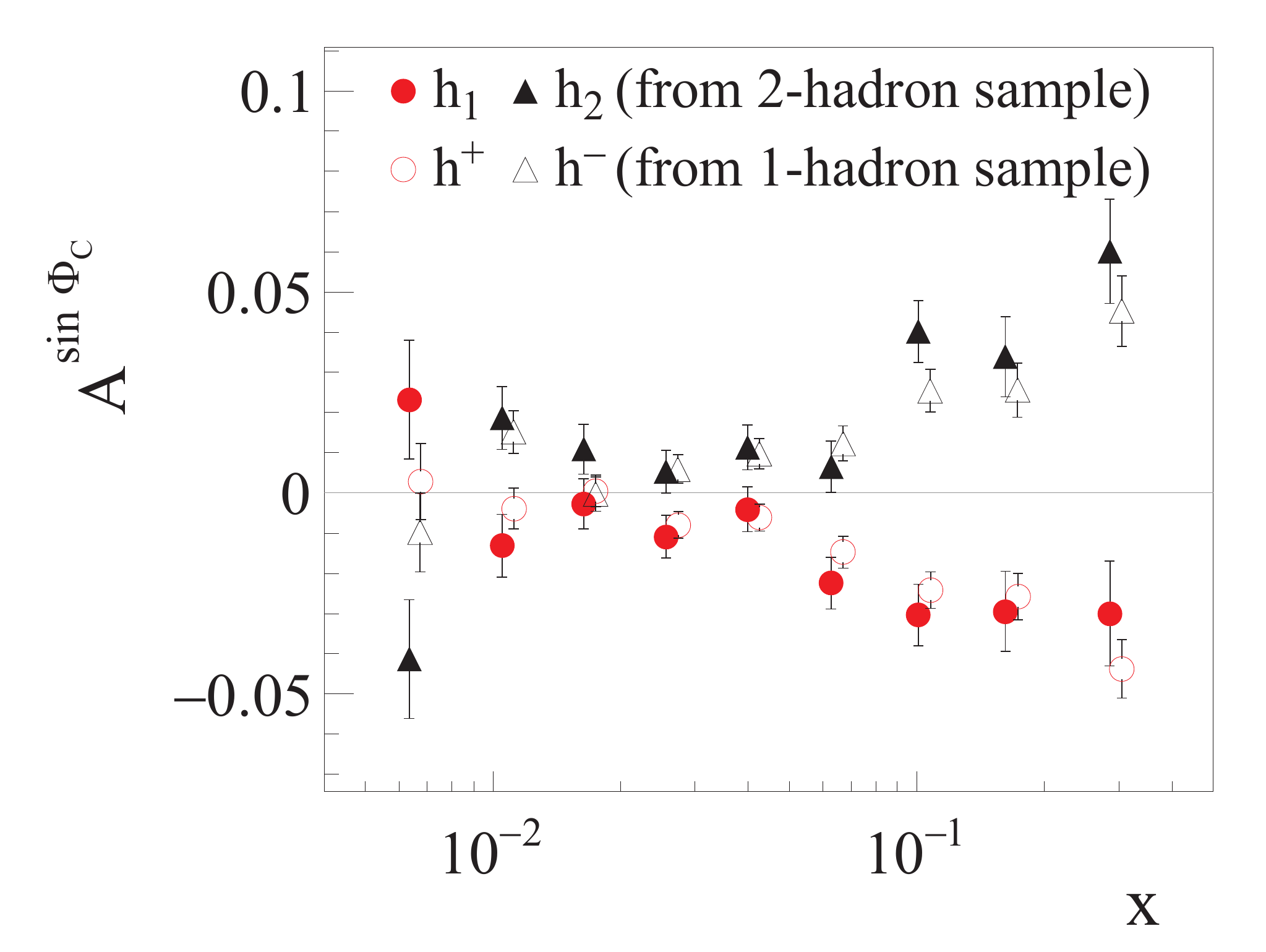}
\caption{(Color online) Comparison of the CL asymmetries for $h_1$ (full red
  circles) and $h_2$ (full black triangles) in $lp \rightarrow l'h^+h^-X$ with
  the standard Collins asymmetries in $lp \rightarrow l'h^{\pm}X$ for $z>0.1$
  (open circles and triangles) measured  as function of $x$.}
\label{fig:cfr1h}
\end{center}
\end{figure}

Within the accuracy of the measurements the CL asymmetries turn out to be the
same as the standard Collins asymmetries (Fig.~\ref{fig:cfr1h} and
Table~\ref{tab:ta1}), implying that the Collins asymmetry does not depend on 
additional observed hadrons in the event.  
As an important result of the first step of this investigation the
2h sample can  be 
used to
study the mirror symmetry and to investigate the interplay between the Collins
single-hadron asymmetry and the di-hadron asymmetry, as described in the
following. 
All the results of the following work are obtained in the kinematical region
$x>0.032$, which is the one where the Collins and the di-hadron asymmetries are
largest.

\begin{table}[b] 
\centering 
\caption{Integrated values of the Collins and CL asymmetries 
  for positive and negative hadrons in the region $x>0.032$. Taking into account
  statistical correlation  the difference is less than a standard deviation for
  both $h_1$ and  $h_2$.
}
\begin{tabular}{|l|r|r|} 
\hline
               & Collins Asymmetry & Collins-like Asymmetry \\ 
\hline 
$h_1$             & $-0.017 \pm 0.002$   &  $-0.018 \pm 0.003 $\\ 
$h_2$             & $0.018 \pm 0.002$   &  $0.020 \pm 0.003$ \\ 
\hline
\end{tabular} 
\label{tab:ta1}
\end{table} 

\section{$\Delta \phi$ dependence of the CL asymmetries of 
positive and negative hadrons}

\begin{figure}[bt]
  \centering
  \includegraphics[width=0.45\textwidth]{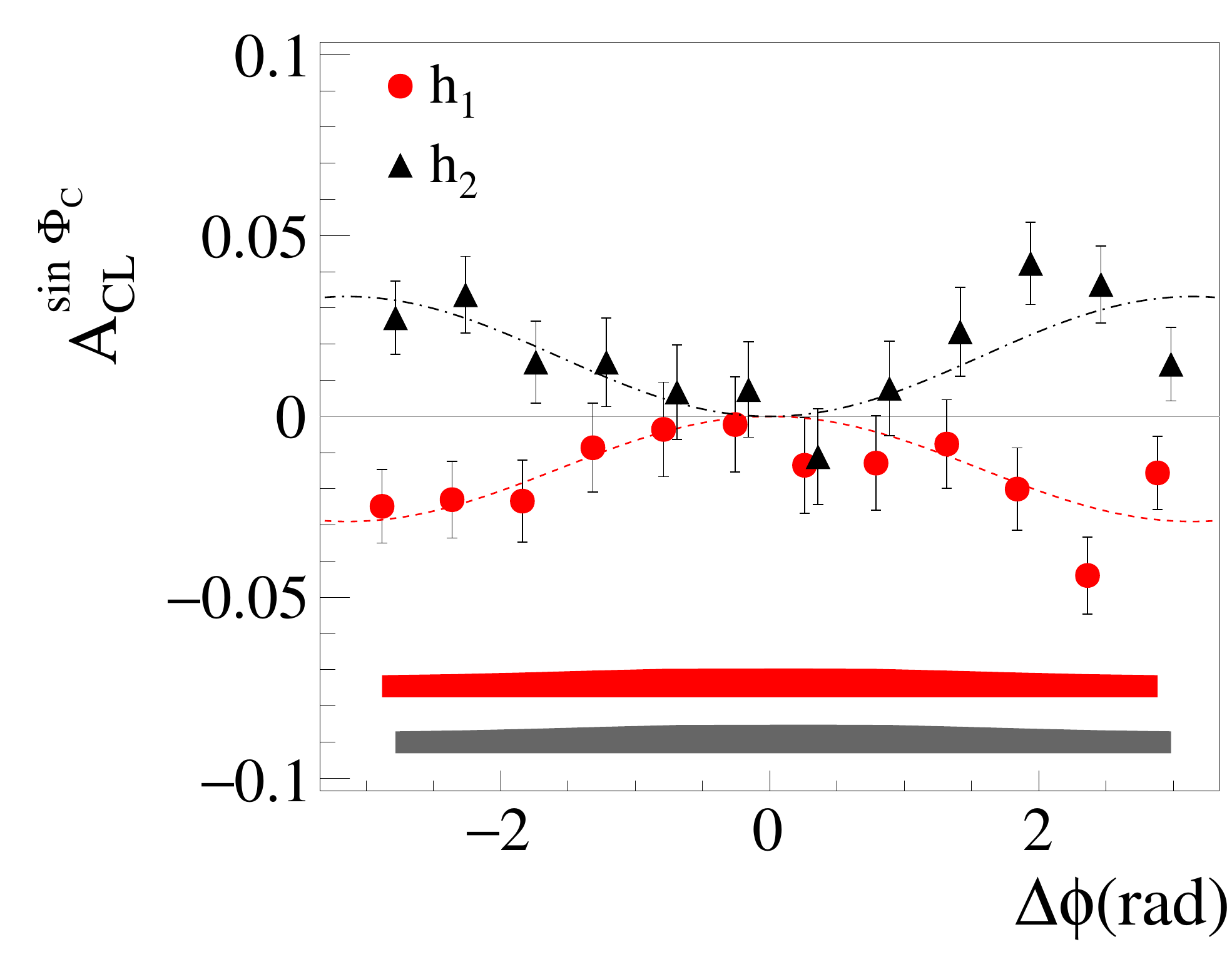}
  \caption{(Color online)
The $A^{\sin\Phi_{C1}}_{CL1}$ (red circles) and the $A^{\sin\Phi_{C2}}_{CL2}$
(black triangles) vs $\Delta\phi$.  Superimposed are the fitting functions
described in the text. 
\label{fig:a1clsin}}
\end{figure}

The azimuthal correlations between $\phi_1$ and $\phi_2$ in transversely
polarized SIDIS had been investigated by measuring the asymmetries as functions
of $| \Delta \phi |$~\cite{Braun:2015baa}, where $\Delta \phi = \phi_1-\phi_2$.
The final results as function of $\Delta \phi$ are shown in
Fig.~\ref{fig:a1clsin}.  The two asymmetries look like even functions of
 $\Delta \phi$, are compatible with zero when $\Delta \phi$ tends to zero, and increase
in magnitude as $\Delta \phi$ increases.
Very much as in Fig.~\ref{fig:cfr1h} the mirror symmetry between positive and
negative hadrons is a striking feature of the data.  The overall picture agrees
with the expectation from the $^3$P$_0$ recursive string fragmentation model of
Refs.~\cite{Artru:1995bh,Artru:2012zz}, which predicts a maximum value for
$\Delta \phi \simeq \pi$. 

\begin{figure}[bt]
  \centering
  \includegraphics[width=0.45\textwidth]{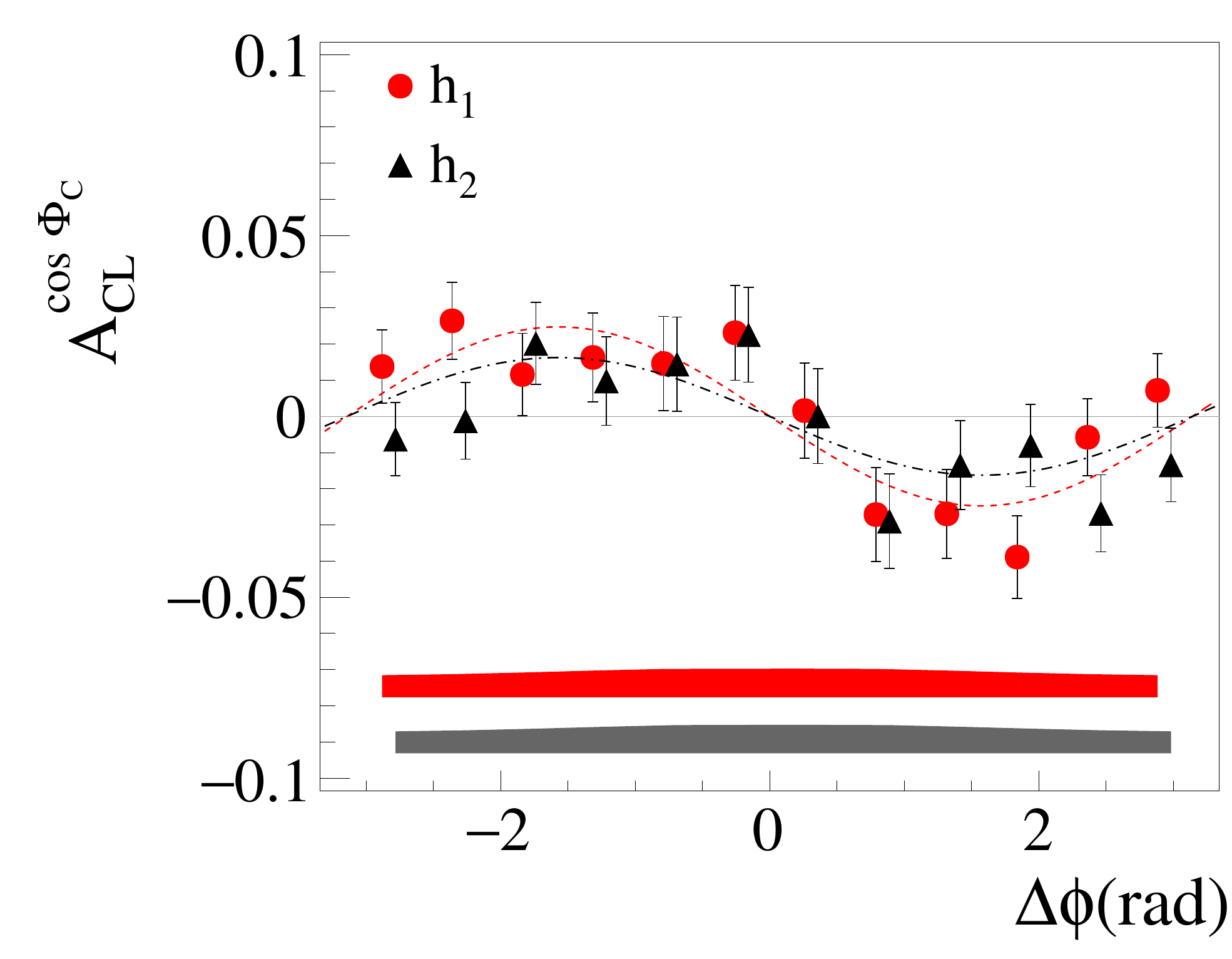}
  \caption{(Color online)
The $A^{\cos\Phi_{C1}}_{CL1}$ (red circles) and the $A^{\cos\Phi_{C2}}_{CL2}$
(black triangles) vs $\Delta\phi$.  Superimposed are the fitting functions
described in the text.
\label{fig:a1clcos}}
\end{figure}

The framework to access the $\Delta \phi$ dependence of CL 
asymmetries was proposed in
Ref.~\cite{Kotzinian:2014uya}. After integration over $x$, $Q^2$, $z_1$, 
$z_2$, $p^2_{T1}$ and $p^2_{T2}$
the cross-section for the SIDIS process $lN \rightarrow l^{'} h^+h^- X$
 can be written as 
\begin{eqnarray}
\label{eq:eq_asy_1}
\frac{d\sigma^{h_1h_2}}{d\phi_1 d\phi_2 d\phi_S}
 &=& \sigma_{U} + S_T \left[
\sigma_{C1} \frac{( \vec{ p}_{T1} \times \vec{q}) \cdot \vec{s'}}{ | \vec{ p}_{T1} \times \vec{q} | \, | \vec{s'} |}
+ \sigma_{C2} \frac{( \vec{ p}_{T2} \times \vec{q}) \cdot \vec{s'}}{ | \vec{ p}_{T2} \times \vec{q} | \, | \vec{s'} |}
\right] \\ \nonumber
 &=& \sigma_{U} + S_T \left[
\sigma_{C1}\sin\Phi_{C1} 
+ \sigma_{C2}\sin\Phi_{C2}
\right], 
\end{eqnarray}
where the unpolarized $\sigma_{U}$ and the polarized $\sigma_{C1}$ and
$\sigma_{C2}$ structure functions (SFs) might depend on $\cos \Delta \phi$.  To
access the azimuthal correlations of the polarized SFs Eq.~(\ref{eq:eq_asy_1})
is rewritten in terms of $\phi_1$ and $\Delta \phi$, or alternatively in terms
of $\phi_2$ and $\Delta \phi$:
\begin{eqnarray}
  \label{eq:eq00}
\frac{d \sigma^{h_1h_2} }{d\phi_1 d \Delta\phi d\phi_S} =  \sigma_U 
 +
S_T 
 \Bigl[
\Bigl( \sigma_{C1}+\sigma_{C2}\cos \Delta\phi \Bigr)
\sin\Phi_{C1} 
  - \sigma_{C2} \sin \Delta\phi \cos\Phi_{C1} \Bigr], 
\nonumber  \\
\frac{d \sigma^{h_1h_2}}{d\phi_2 d \Delta\phi d\phi_S}   =  \sigma_U
 +
S_T 
\Bigl[
\Bigl(  \sigma_{C2}+\sigma_{C1}\cos \Delta\phi \Bigr)
\sin\Phi_{C2}  
   + \sigma_{C1} \sin \Delta\phi  \cos\Phi_{C2} \Bigr] . 
\end{eqnarray}
With the change of variables above a new modulation, of the type $\cos
\Phi_{C1,2}$, appears in the cross section, which can then be rewritten in terms
of the sine and cosine modulations of the Collins angle of either the positive
or the negative hadron.  The explicit expressions for the four asymmetries are:
\begin{eqnarray}
  \label{eq:eq1}
  A_{CL1}^{\sin\Phi_{C1}} = \frac{1}{D_{NN}}
 \frac{\sigma_{C1}
    +\sigma_{C2} \cos \Delta \phi}{\sigma_U} \, , \; \; \; \;
  A_{CL1}^{\cos\Phi_{C1}} = - \frac{1}{D_{NN}}
  \frac{\sigma_{C2}
    \sin \Delta \phi}{\sigma_U} \, ,  \nonumber \\
  A_{CL2}^{\sin\Phi_{C2}} = \frac{1}{D_{NN}}
\frac{\sigma_{C2}
    +\sigma_{C1} \cos \Delta \phi}{\sigma_U} \, , \; \; \; \;
  A_{CL2}^{\cos\Phi_{C2}} = \frac{1}{D_{NN}}
\frac{\sigma_{C1}
    \sin \Delta \phi}{\sigma_U} \, ,
\end{eqnarray}
where $D_{NN}$ is the mean
transverse-spin-transfer coefficient, equal to 0.87 for these data.  
Figure~\ref{fig:a1clcos} shows the measured
values of the new asymmetries $A_{CL1,2}^{\cos\Phi_{C1,2}}$.  
It is the first time that they are measured.
They have rather
similar values for positive and negative hadrons,  seem to be 
odd functions of $\Delta\phi$, and average  to
zero when integrating over $\Delta\phi$.  
Note that the data are
in very good agreement with Eq.~(\ref{eq:eq1}) if $(\sigma_{C1}/\sigma_U) = -
(\sigma_{C2}/\sigma_U)=\mathrm{const.}$. 

\begin{figure}[bt]
  \centering
  \includegraphics[width=0.45\textwidth]{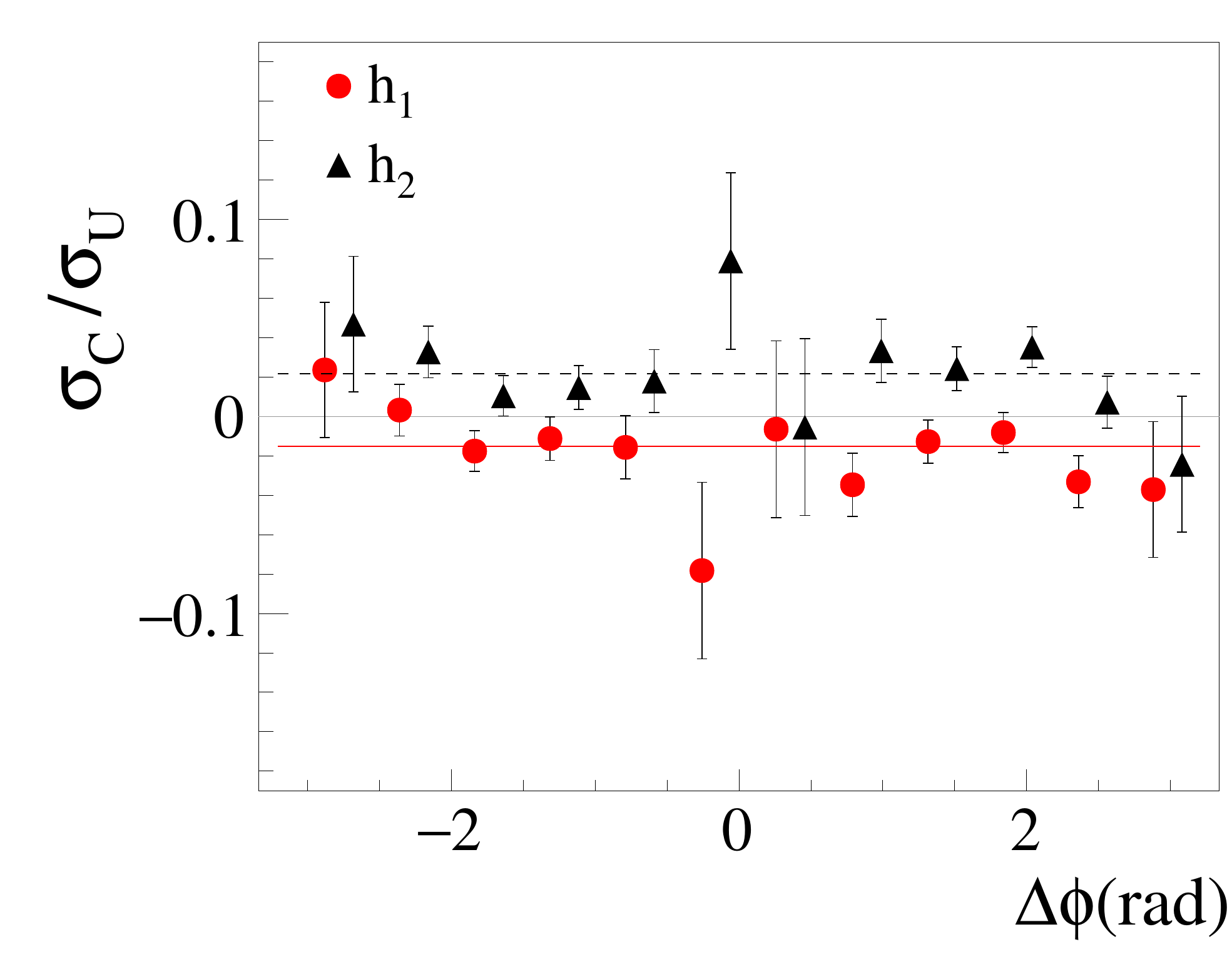}
\caption{(Color online) The measured values of the ratios $\sigma_{C1} / \sigma_{U}$ and
  $\sigma_{C2} / \sigma_{U}$.  The lines give the fitted mean values, $-0.015
  \pm 0.004$ for $h_1$ and $0.022 \pm 0.004$ for $h_2$.
 \label{fig:rcs}}
\end{figure}

The quantities $\sigma_{C1} / \sigma_{U}$ and $\sigma_{C2} / \sigma_{U}$, which
in principle can still be functions of $\Delta \phi$, can be obtained from the
measured asymmetries using
\begin{eqnarray}
\label{eq:eq3}
\frac{\sigma_{C1}}{\sigma_U}
&=& D_{NN} \left[
A_{CL1}^{\sin \Phi_{C1}}  + 
A_{CL1}^{\cos \Phi_{C1}} \cot \Delta \phi \right] \, ,
 \nonumber \\
\frac{\sigma_{C2}}{\sigma_U}
&=& D_{NN} \left[
A_{CL2}^{\sin\Phi_{C2}}  -
A_{CL2}^{\cos \Phi_{C2}} \cot \Delta \phi \right].
\end{eqnarray}
The values of the ratios $\sigma_{C1} /\sigma_{U}$ and $\sigma_{C2} /\sigma_{U}$
extracted from the measured asymmetries are given in Fig.~\ref{fig:rcs}.  Within
the statistical uncertainty they are constant, hinting at similar azimuthal
correlations in polarized and unpolarized SFs.  Moreover, they are almost equal
in absolute value and of opposite sign.
Assuming $(\sigma_{C1}/\sigma_U) = - (\sigma_{C2}/\sigma_U) = \mathrm{const.}$,
the measured asymmetries can be fitted with the simple functions $\pm a
(1-\cos\Delta\phi)$ in the case of the sine asymmetries, and $a \sin\Delta\phi$
for the cosine asymmetries.  The results of the fits for positive (negative)
hadrons are the dashed red (dot-dashed black) curves shown in
Fig.~\ref{fig:a1clsin} and \ref{fig:a1clcos}.  The agreement with the
measurements is very good and the four values for the constants $a$ are well
compatible, as can be seen in Table~\ref{tab:fits}.
\begin{table}[b] 
\centering 
\caption{Fitted values of the $a$ parameter for the two CL asymmetries
  for positive and negative hadrons.  }
\begin{tabular}{|l|c|c|} 
\hline
               & $A^{\sin\Phi_{C}}_{CL}$ & $A^{\cos\Phi_{C}}_{CL}$ \\ 
\hline 
$h_1$             & 0.014 $\pm$ 0.003   &  0.025 $\pm$ 0.005 \\ 
$h_2$             & 0.016 $\pm$ 0.003   &  0.017 $\pm$ 0.005 \\ 
\hline
\end{tabular} 
\label{tab:fits}
\end{table} 

As a conclusion of this 
second step of the investigation,
the $h_1$ and $h_2$ CL asymmetries as
functions of $\Delta \phi$ agree with the expectation from the $^3$P$_0$
recursive string fragmentation model and with the calculations of the $\Delta
\phi$ dependence obtained in Ref.~\cite{Kotzinian:2014uya}.  As in the
one-hadron sample a mirror symmetry for the positive and negative hadron sine
asymmetries is observed in the 2h sample, which is a consequence of the
experimentally established relation $\sigma_{C1} = - \sigma_{C2} $.

These results allow to derive a quantitative relation between the $h_1$ and
$h_2$ CL asymmetries and the di-hadron asymmetry, as described in the following.

\section{Comparison of  CL and  di-hadron 
asymmetries}

The third and last step of this investigation has been the formal 
derivation of a connection between the
CL and the di-hadron asymmetries and the comparison with the experimental
data.
In the standard analysis, the $\Delta\phi$ integrated di-hadron asymmetry is
measured from the amplitude of the sine modulation of the angle
$\Phi_{RS}=\phi_R+\phi_S-\pi$, where $\phi_R$ is the azimuthal angle of the
relative hadron momentum 
$\vec{ R} =\left[ z_2 \vec{ p}_1-z_1 \vec{ p}_2 \right]
/ \left[ z_1 + z_2 \right]=:\xi_2 \vec{ p}_1 -
\xi_1 \vec{ p}_2$.
In the present analysis, the azimuthal angle
$\phi_{2h}$ of the vector $\vec{R}_N=\hat{p}_{T1} - \hat{p}_{T2}$ is evaluated for each pair of
oppositely charged hadrons, with the hat indicating unit vectors.  
As discussed
in Ref.~\cite{Adolph:2014fjw}, the  azimuthal angle $\phi_R$
is strongly correlated with 
$\phi_{2h}=[\phi_{1}+\phi_{2}+ \pi \sgn(\Delta \phi)]/2$,  where $\sgn$ is the
signum function.  
Also,
introducing the angle 
$\Phi_{2h,S}=\phi_{2h}+\phi_{S}-\pi$, 
which is a kind of
mean of the Collins angle of the positive and negative hadrons after correcting
for a $\pi$ phase difference, it was shown~\cite{Adolph:2014fjw} that the
di-hadron asymmetry measured from the amplitude of $\sin \Phi_{2h,S}$ is
essentially identical to the standard di-hadron asymmetry.  
In order to establish a connection between the di-hadron asymmetry
and the CL asymmetries
$\Phi_{2h,S}$ will be used rather than $\Phi_{RS}$ in the following.  
Starting from the general
expression for the cross section given in Eq.~(\ref{eq:eq_asy_1}), changing
variables from $\phi_1 $ and $\phi_2$ to $\Delta \phi$ and $\phi_{2h}$, and
using the relations $\sin \Phi_{2h,S} = ( \hat{R}_N \times \hat{q}) \cdot
\hat{s'}$ and $\cos \Phi_{2h,S} = - \sgn(\Delta\phi) ( \hat{P}_N \times \hat{q})
\cdot \hat{s'}$, where $\vec{P}_N=\hat{p}_{T1} + \hat{p}_{T2}$,
Eq.~(\ref{eq:eq_asy_1}) can be rewritten as:
\begin{eqnarray}
\label{sigma-phi_2h-phi_R_N}
\frac{d\sigma^{h_1 h_2}}{d\Delta\phi\, d\phi_{2h} \,d\phi_S}&=&
\sigma_{U}+S_T\frac{1}{2}\big[\big(\sigma_{C1}-\sigma_{C2}\big)\sqrt{2(1-\cos\Delta\phi)}\sin\Phi_{2h,S} \nonumber \\
&&-\sgn(\Delta\phi)\big(\sigma_{C1}+\sigma_{C2}\big)\sqrt{2(1+\cos\Delta\phi)}
\cos\Phi_{2h,S}\, \big],
 \label{eq:s2hc}
\end{eqnarray}
which simplifies to
 \begin{eqnarray}
 \frac{d\sigma^{h_1h_2}}{d\phi_{2h} d \Delta\phi d\phi_S}
 &=& \sigma_{U}  + S_T \cdot \sigma_{C1} \cdot
 \sqrt{2(1-\cos \Delta\phi )}\cdot \sin\Phi_{2h,S}
 \label{eq:s2h}
 \end{eqnarray}
using the experimental result $\sigma_{C2} = - \sigma_{C1}$.  This last
cross-section implies a sine modulation with amplitude
\begin{eqnarray}
A^{\sin\Phi_{2h,S}}_{CL2h} = \frac{1}{D_{NN}} 
\frac{
  \sigma_{C1}}{\sigma_{U}} \cdot
\sqrt{2(1-\cos \Delta\phi)} .
\label{eq:2h2}
\end{eqnarray}
  \begin{figure}[tb]
  \centering
  \includegraphics[width=0.45\textwidth]{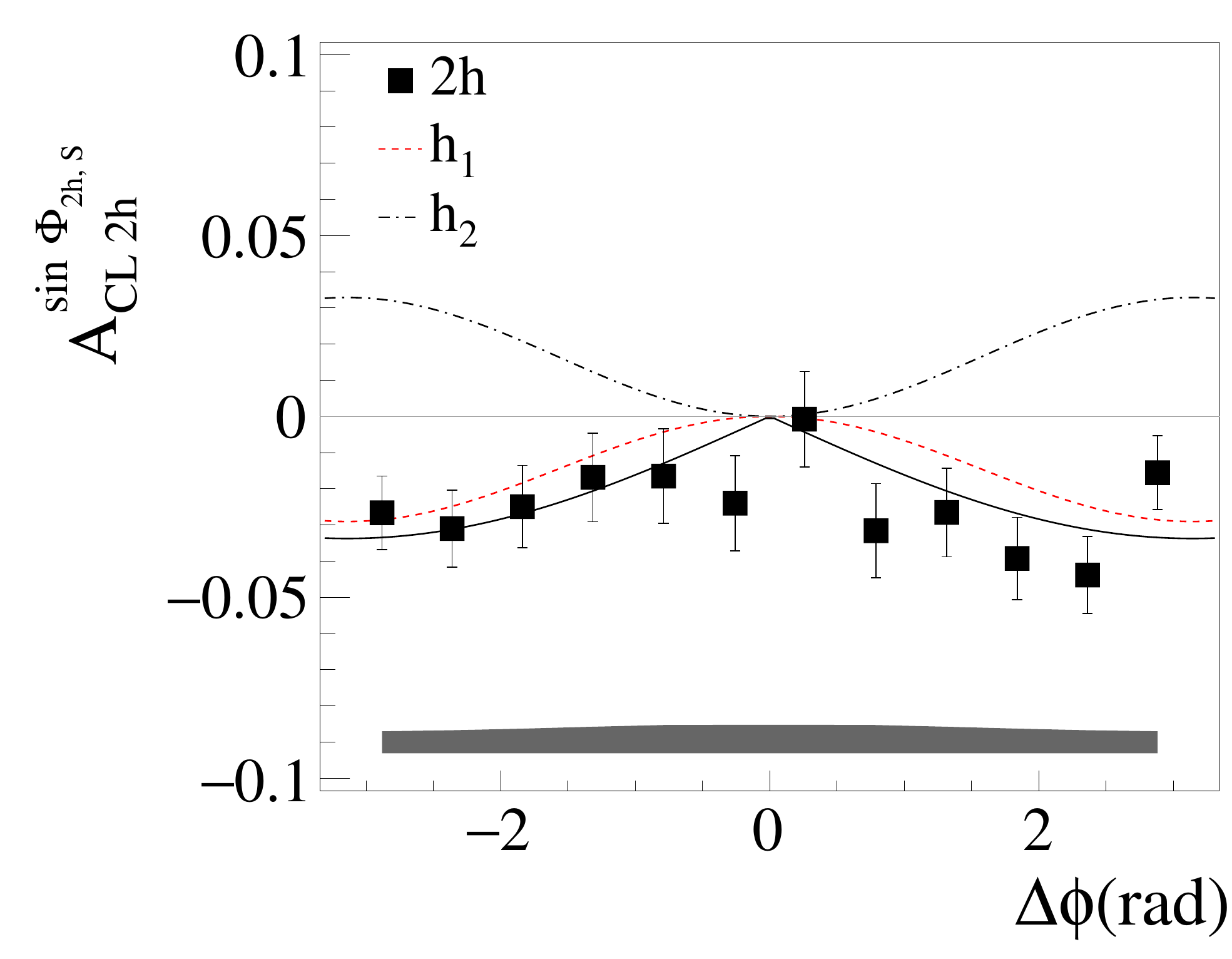}
  \caption{(Color online) 
$A^{\sin\Phi_{2h,S}}_{CL2h}$ vs $\Delta\phi$ and the corresponding fit (black
    full curve).  The dashed red and dot-dashed black curves are the fits to
    $A^{\sin\Phi_{C1}}_{CL1}$ and $A^{\sin\Phi_{C2}}_{CL2}$ from
    Fig.~\ref{fig:a1clsin}. 
\label{fig:a12hsin}}
  \end{figure}
At variance with the single hadron case, no $A^{\cos\Phi_{2h,S}}_{CL2h}$
asymmetry is present in Eq.~(\ref{eq:s2h}) and the measured values are indeed
compatible with zero.  In Figure~\ref{fig:a12hsin} the
$A^{\sin\Phi_{2h,S}}_{CL2h}$ asymmetry is shown together with the curve
$c\sqrt{2(1-\cos \Delta\phi)}$ with $c= -0.017 \pm 0.002$ (black solid line) as
obtained by the fit.  The dashed red and dot-dashed black curves are the fitted
curves $a (1-\cos \Delta\phi)$ of Fig.~\ref{fig:a1clsin}.  As can be seen
the fit is good, and the value of $c$ is well compatible with the
corresponding values of Table~\ref{tab:fits}, in agreement with the fact that 
$\sigma_{Ci} / \sigma_{U}$ is the same for the three asymmetries.
Evaluating the ratio of the integrals of the di-hadron amplitudes over the
one-hadron amplitudes one gets a value of $1.4 \pm 0.2$ which agrees with the
value $4/\pi$ evaluated from Eqs.~(\ref{eq:2h2}) and~(\ref{eq:eq1}) and with
our original observation that the di-hadron asymmetry is somewhat larger than
the Collins asymmetry for positive hadrons.

\section{Conclusions}

We have shown that in SIDIS hadron-pair production the $x$-dependent
Collins-like single hadron asymmetries of the positive and
negative hadrons are well compatible with the standard Collins 
asymmetries and are mirror symmetric.
Also, the Collins-like asymmetries exhibit a $\pm a(1 - \cos \Delta \phi)$
dependence on $\Delta \phi$, which we have derived from the general expression
for the two-hadron cross-section and is a consequence 
of the experimentally 
verified similar $\Delta \phi$
dependence of the unpolarized and polarized structure functions 
and the mirror symmetry of the last ones.

Most important, for the first time it has been shown that the 
amplitude of the di-hadron asymmetry as a function of $\Delta \phi$
has a very simple relation to that of the single hadron asymmetries 
in the 2h sample, namely it can be
written as a $a\sqrt{2(1 - \cos \Delta\phi)}$, where the constant $a$ 
is the same as that which appears in the expressions
for the Collins-like asymmetries. 
After integration on $\Delta \phi$, the di-hadron asymmetry has to be 
larger than
the single hadron asymmetries by a factor $4/\pi$, in good agreement 
with the measured values.

In conclusion, we have shown that the integrated values of Collins 
asymmetries in the 1h sample are the
same as the Collins-like asymmetries of 2h sample which in turn are 
related with the integrated values of di-hadron
asymmetry. This gives indication that both the single hadron and 
di-hadron transverse-spin dependent
fragmentation functions are driven by the same elementary mechanism. 
As a consequence of this important
conclusion we can add that the extraction of transversity distribution 
using the di-hadron asymmetry
in SIDIS does not represent an independent measurement with respect 
to the extractions which are based
on the Collins asymmetry.

\end{document}